\documentclass[10pt,journal,a4paper]{IEEEtran}
\usepackage{amsmath,epsfig}
\usepackage{amssymb}
\usepackage{times}
\usepackage{color}
\usepackage{graphicx}
\usepackage{theorem}
\usepackage{cite}
\usepackage{euscript}
\usepackage{pifont}
\def\ben{\begin{enumerate}}
\def\beq{\begin{equation}}
\def\beqa{\begin{eqnarray}}
\def\bit{\begin{itemize}}
\def\een{\end{enumerate}}
\def\eeq{\end{equation}}
\def\eeqa{\end{eqnarray}}
\def\eit{\end{itemize}}

\def\insertfig#1#2#3#4#5{
\begin{figure}[#1]
\centering\includegraphics[width=#2\columnwidth,clip=]{#3.eps}
\caption{#4}\label{#5}\end{figure}}
\title{Multiuser detection in a dynamic environment \\
Part I: User identification and data detection}
\author{Ezio~Biglieri,~\IEEEmembership{Fellow,~IEEE,
} and~Marco~Lops,~\IEEEmembership{Senior Member,~IEEE}%
\thanks{Manuscript received XX, XXXX; revised
YY, YYYY. }%
\thanks{Ezio Biglieri is with Universitat Pompeu Fabra, Barcelona, Spain. email: {\tt
ezio.biglieri@upf.edu} This work was supported by the STREP
project No.\ IST-026905 (MASCOT) within the 6th framework program
of the European Commission, and by the Spanish Ministery of
Education and Science under Project TEC2006-01428/TCM.}
\thanks{Marco Lops is with DAEIMI, Universit\`a di Cassino, Italy.
email: {\tt lops@unicas.it}}}

\begin{document}
\maketitle

\begin{abstract}
 In random-access communication systems,  the
number of active users varies with time, and has considerable
bearing on receiver's performance. Thus, techniques aimed at
identifying not only the information transmitted, but also that
number, play a central role in those systems. An example of
application of these techniques can be found in multiuser
detection (MUD). In typical MUD analyses, receivers are based on
the assumption that the number of active users is constant and
known at the receiver, and coincides with the maximum number of
users entitled to access the system. This assumption is often
overly pessimistic, since many users might be inactive at any
given time, and detection under the assumption of a number of
users larger than the real one may impair performance.

The main goal of this paper is to introduce a general approach to
the problem of identifying active users and estimating their
parameters and data in a random-access system where users are
continuously entering and leaving the system. The tool whose use
we advocate is Random-Set Theory: applying this, we derive optimum
receivers in an environment where the set of transmitters
comprises an unknown number of elements. In addition, we can
derive Bayesian-filter equations which describe the evolution with
time of the a posteriori probability density of the unknown user
parameters, and use this density to derive optimum detectors. In
this paper we restrict ourselves to interferer identification and
data detection, while in a companion paper we shall examine the
more complex problem of estimating users' parameters.
\end{abstract}
\section{Introduction}
{
 In typical random-access communication systems,
the number of active users, their location, as well as the
parameters that characterize their channel state, vary with time.
Thus, techniques aimed at identifying not only the data
transmitted, but also the user parameters, play a central role in
analysis and design of those systems. Examples of application of
these techniques can be found in multiuser detection (MUD),
spatial multiplex schemes, and ad hoc networks. }

{
 In MUD, it has long been recognized that one of the
important issues is that the set of active users at any time may
not be known to the receiver. For conventional (matched-filter)
receivers, this dearth of information does not affect performance,
while for other receivers the simplistic assumption that all users
are active will cause significant degradation.~\footnote{See,
e.g.,~\cite{chentong,mitpoo1,wuchen}. As an example, if a
decorrelator detector~\cite[Chapter 5]{verdubook} does more
nulling than needed, its performance is impaired.
Ref.~\cite{honpoo} describes a case  where a multiuser detector
suffers from catastrophic error if a  new user becomes active.} In
addition, certain  detectors based on interference cancellation
must know the  strongest active users in order to perform
satisfactorily~\cite{halbra}. Moreover, as observed in~\cite{xu},
``identification of active users will help the system to promptly
process requests and efficiently allocate channels. In such a way,
system capacity can be increased.''

The problem of detecting active users in a multiuser system has
been addressed by several authors in a CDMA
context~\cite{buzdemlop,halbra,mitpoo1,mitpoo2,oskpoo,wuchen}.
Typically, the resulting multiuser receiver is a combination of
two separate modules, namely, the active-user identifier and the
multiuser detector. The treatment in~\cite{halbra,mitpoo1,mitpoo2}
focuses on the problem of detecting a single user entering or
leaving the system.  Ref.~\cite{wuchen} advocates a subspace-based
method (MUSIC algorithm) for identifying the active users, under
the assumption that the receiver knows the pool of all possible
spreading codes that may be used in transmission.
In~\cite{linlim}, the active-user-identification algorithm is
subspace-based, as in~\cite{wuchen}; the receiver is not
interested in decoding all active users, but only those
transmitting a message to it. Ref.~\cite{xu} addresses the problem
of estimating the number of active users when synchronous and
asynchronous users co-exist in the system.

Further performance improvement can be expected if the receiver
can exploit a priori information about users entering and exiting
the system. The knowledge of a traffic model is exploited
in~\cite{chentong}, whose authors use it to improve the detection
of active users. They model bursty traffic for an individual
source as a two-state Markov chain.

Other systems in which user identification is necessary include
spatial multiplexing schemes, where the total system throughput
can be optimized by properly selecting a subset of users to which
the power is allocated~\cite{madanskha,yurhe}. Thus, optimum
power-control strategy requires the identification of this best
set of users. In ad hoc networks, optimal transmission strategies
require the identification and localization of active nodes in the
neighborhood of the transmitter.
}

\subsection{In this paper\dots}
{
 In this paper, and in its companion~\cite{angbiglop},
we examine the general problem of detecting data and parameters of
a set of users whose number is unknown. Unlike previous work done
in this area, which advocates a two-stage receiver, we focus on
optimal design of a receiver which estimates simultaneously the
number of users and their parameters and data.
 The receiver performance can be enhanced if one uses the a
priori information offered by a traffic model, i.e., a dynamic
model for the number of active users and for their parameters. In
fact, information from the past history of the parameters may
bring a considerable amount of extra information if their changes
are not overly abrupt (for example, if the number of active users
does not change considerably from frame to frame).

{\color{black} Having to deal with random sets, i.e., with sets
comprising a random number of random vectors (those including what
is unknown about each user), a tool that can be used is {\em
Random Set Theory} (RST: see the Appendix and the references
therein). RST, recently  applied in the context of multitarget
tracking and identification (see,
e.g.,~\cite{goodman,mahlerreport,mahler2003,mahler2004,mahlerbook,vihola,vosindou})
is based on a probability theory of finite sets that exhibit
randomness not only in each element, but also in the number of
elements.  RST (and in particular its formulation, referred to as
Finite Set Statistics, or
FISST~\cite{mahlerreport,mahler2003,mahler2004}, specifically
tailored to problems in whose class lie those we are considering
in this paper) develops concepts which are not part of
conventional probability theory. In fact, a central point in FISST
is the generation of ``densities'' which are not the usual
Radon-Nikod\'ym derivatives of probability measures, but rather
``set derivatives'' of nonadditive ``belief functions.'' On the
other hand, these densities, which capture what is known about
measurement state space, users state space, and users dynamics,
can be derived in a rather straightforward way from the system
model by using the FISST toolbox. RST is a tool that has
considerable generality and flexibility, is consistent with
engineering intuition, and is easy to use.\footnote{\color {black}
Advocating the use of RST to solve the problem at hand, we do not
imply that it is the only tool that can be used. Actually,
standard probability theory could be applied to achieve the same
results, although, we argue, in a less elegant and concise way.
The companion paper~\cite{angbiglop}, which deals with estimates
of random sets of continuous parameters, should be even more
convincing about the usefulness of RST.}}


To illustrate application of RST to random-access communication,
we focus on MUD problems, and derive Bayesian-filter equations
describing the evolution with time of the a posteriori probability
density of the unknown user parameters and data. Specifically,
here we restrict ourselves to interferer identification and data
detection, while in a companion paper~\cite{angbiglop} we examine
the problem of estimating users' parameters as well. We hasten to
claim that the applications considered here do not exhaust the
potential of RST for the analysis of random-access system: thus,
many of the simplifying assumptions are not made because more
realistic models cannot be dealt with using our theory, but rather
because we do not want to muddle the intrinsic simplicity of the
RST tool with marginal details. This paper is organized as
follows. Section~\ref{model} describes the channel model, while
Section~\ref{detection} states the MUD problem in the context of
RST. Section~\ref{example} describes an application to CDMA, while
Section~\ref{results} provides some numerical results illustrating
the theory.

 }
\section{Channel model and statement of the problem}\label{model}
We assume $K+1$ users transmitting digital data over a common
{
random-access} channel. Let  $s({\bf x}_t^{(0)})$
denote the signal transmitted by the reference user at discrete
time $t$, $t=1, 2,\ldots$,
 and $s({\bf x}_t^{(i)})$, $i=1, \ldots, K$, the signals
 that may be transmitted at the same time by $K$ interferers.
 Each signal has in it a number of known parameters, reflected by the
 deterministic function $s(\, \cdot \,)$, and a number of random parameters, summarized
 by ${\bf x}_t^{(i)}$. The index $i$ reflects the identity of the
 user, and is typically associated with its signature. The
 observed signal at time $t$ is a
 sum of $s({\bf x}_t^{(0)})$, of the signals generated by the users active at time $t$,
 which are in a random number, and
 of stationary random noise ${\bf z}_t$. We write
  \beq\label{formulauno}
  {\bf y}_t = s({\bf x}^{(0)}_t) + \sum_{{\bf x}_t^{(i)}\in {\bf X}_t}
  s({\bf x}_t^{(i)}) +{\bf z}_t
  \eeq
 where ${\bf X}_t$ is a random set, encapsulating what is unknown about the active users.
The notations of~(\ref{formulauno})
 implicitly assume that user $0$ is active with probability $1$
 and its parameters (but not its data) are known (this restriction can be easily removed).

To motivate the development presented in this paper, and in
particular our use of RST, we proceed to formulate the general
problem through three intermediate steps. Specifically, we examine
three scenarios of increasing complexity, under the assumptions
that the users' parameters are all known, that the number of
interferers is random and unknown to the receiver, and that we are
interested in detecting the data transmitted by the reference
user:
\begin{dingautolist}{192}
\item
{\em The receiver has no information about the a priori
probabilities that the individual interferers are active.} Two
options we may consider here are maximum-likelihood (ML) detection
of the reference user's data under the assumption that all
potential interferers are active, or joint ML detection of the
number of active interferers and of the reference user's data.
Consider binary transmission for simplicity. In the first case,
detection implies choosing among $2\times 2^K$ hypotheses. In the
second case, the choice is among $2\times 3^K$ hypotheses, as
every interferer may transmit one between two binary symbols, or
be inactive. The difference in performance between the two
situations is illustrated in Fig.~\ref{new_figure_1}, which
compares the two detectors described above.
\insertfig{t}{0.8}{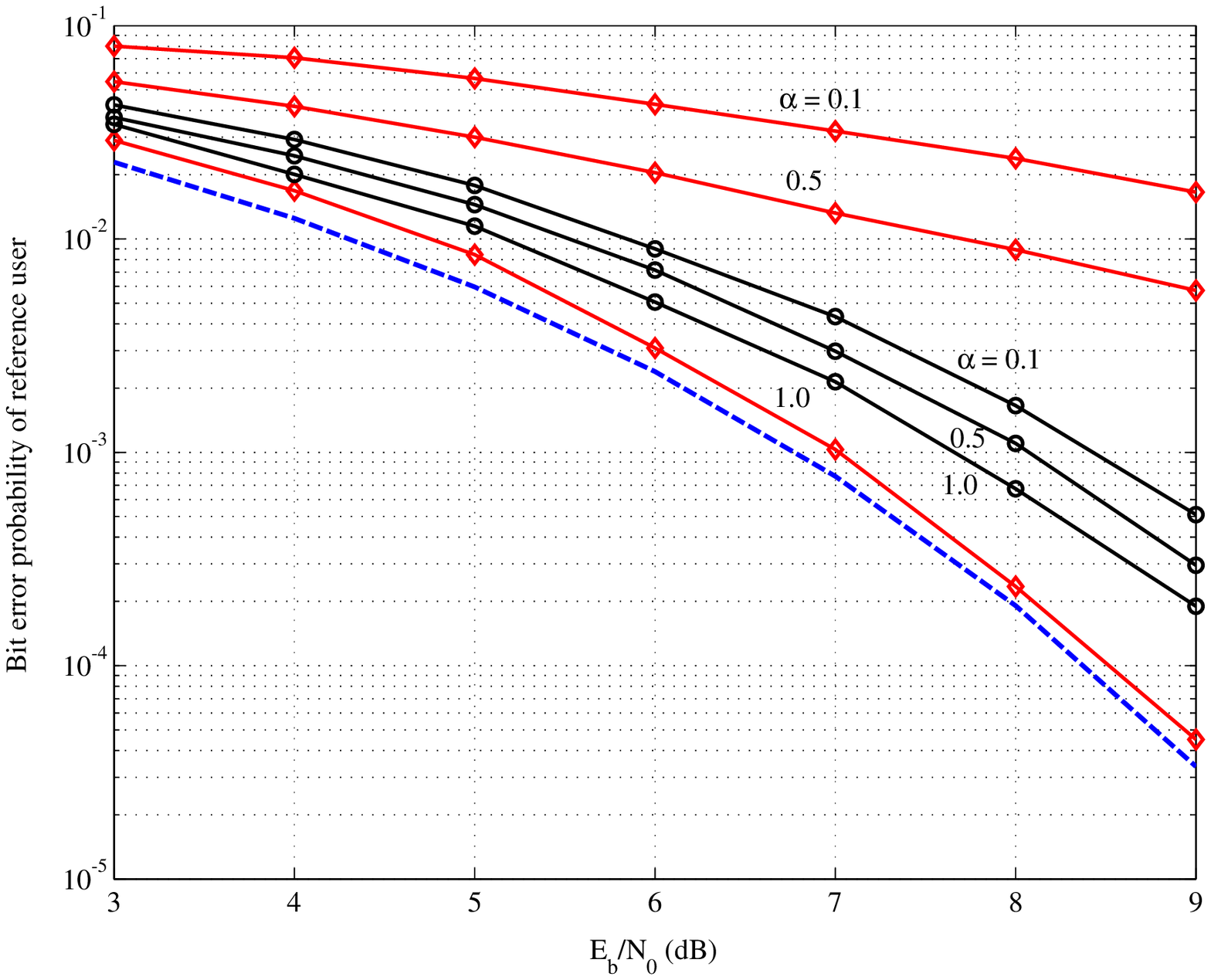}{\sl Bit error probability of the
reference user in a multiuser system with $2$ interferers,
independently active with probability $\alpha$. Lines with diamond
markers: Classic multiuser ML detection, assuming that all users
are active. Lines with circle markers: ML detection of data and
interferers number. Dashed curve: Single-user
bound.}{new_figure_1}
The ordinate shows the bit error probability of the reference user
in a multiuser system with $2$ independent interferers
transmitting binary antipodal signals over an additive white
Gaussian noise (AWGN) channel with the same a priori probability
of activity $\alpha$, spreading-sequences consisting of Kasami
sequences with length $15$~\cite[p.\ 240]{ipatov}, and perfect
power control. The single-user bound is also shown as a reference.
It is seen that RST yields a detector much more robust than
classic MUD to variations in the users activity factor. We also
observe that classic MUD can outperform RST for high values of
$\alpha$, as this situation corresponds to its having reliable
side information about the number of active users.
\item
{\em The receiver knows the a priori probabilities that the
individual interferers are active.} System performance can be
further improved if the receiver is able to exploit additional
side information in the form of a priori probabilities of user
activity. By assuming that the activity factor $\alpha$ is known,
maximum a posteriori (MAP) detection yields the results shown in
Fig.~\ref{new_figure_2}.
\insertfig{t}{0.8}{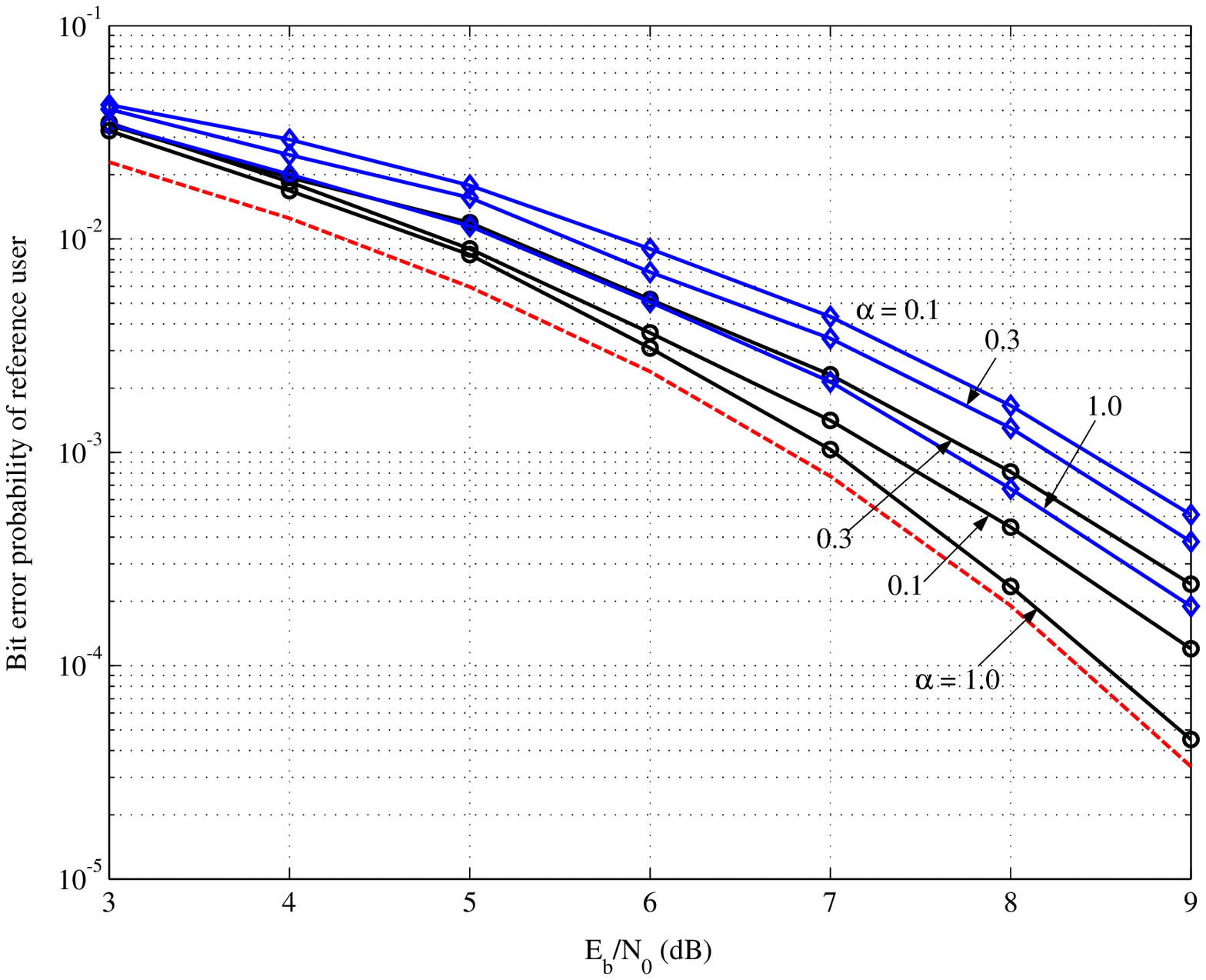}{\sl Bit error probability of the
reference user in a multiuser system with $2$ interferers,
independently active with probability $\alpha$. Lines with diamond
markers: ML detection of active users and data. Line with circle
markers: MAP detection using a priori knowledge of the value of
$\alpha$. Dashed curve: Single-user bound.}{new_figure_2}
\item
{\em The receiver has a dynamic model of users' activity.} The
receiver performance can be further improved by using additional
information about the interferers, in the form of a model of their
dynamic behavior. This information can be generated once a model
of users' mobility is available.
\end{dingautolist}

Observe again that the information carried by the interferers is
contained in the set
\[
{\bf X}_t = \{ {\bf x}_t^{(1)} , \ldots, {\bf x}_t^{(k)} \}
\]
whose elements are random vectors, and $k$ is itself a random
integer. RST develops a probability theory on random sets of this
form, which are modeled as single entities.  Roughly speaking, a
random set is a map ${\bf X}$ between a sample space and a family
of subsets of
 a space ${\mathbb S}$. This is the space of the unknown data and parameters
 of the active interferers. For example,
 if everything about the interferers is known, except for their
 number and identity, then $\bf X$ takes values in the power set
 $2^{\mathbb K}$, where ${\mathbb K}\triangleq \{ 1, \ldots , K \}$.
We may also consider a
situation in which one or more parameters (the interferer power,
etc.) are also unknown in addition to the interferers'
 number and identities, while the transmitted data are known (for example, in a training phase).
 In mathematical terms, $\mathbb S$ is generally a {\em hybrid space} ${\mathbb
 S}\triangleq {\mathbb R}^d \times U$, with $U$
 a finite discrete set, and $d$ the number of parameters to be estimated for each user.
In the remainder of this paper we shall restrict ourselves to the
case $d=0$, and leave to a companion paper~\cite{angbiglop} the
discussion of the case $d\neq 0$.

 With channel model~(\ref{formulauno}), the receiver detects only a superposition
 of interfering signals. Thus, the random set describing the
 receiver, denoted ${\bf Y}_t$, is the singleton $\{ {\bf y}_t
 \}$, where ${\bf y}_t$ has conditional
 probability density function
 \beq
 f_{{\bf Y}_t|{\bf X}_t}(  {\bf y}_t   \mid {\bf B}) = f_{\bf z}({\bf y}_t - \sigma({\bf B}))
 \eeq
 where ${\bf B}= \{ {\bf b}_1, \ldots, {\bf b}_k \}$ is a realization of ${\bf X}_t$,
 that is, a realization of a random set of users and their
 parameters\slash data. Moreover,
 $f_{\bf z}(\, \cdot  \,)$ is the probability density function (pdf) of the
 additive noise, and
 \beq
 \sigma({\bf B}) \triangleq \sum_{{\bf b}_i\in {\bf B}}
  s({\bf b}_i)
 \eeq

\subsection{Defining estimators}
Development of estimators with our model must take into account
the peculiarities of RST. {\color{black} For example, expectations
cannot be defined, because there is no notion of set addition, and
hence estimators based on a posteriori expectations do not
exist~(this point is discussed thoroughly and eloquently
in~\cite{mahlerreport}).} A possible estimator maximizes the a
posteriori probability (APP) of ${\bf X}_t$ given ${\bf y}_{1:T}$,
the latter denoting the whole set of observations corresponding to
a data frame transmitted from $t=1$ to $t=T$. Another possibility
is to restrict oneself to a {\em causal} estimator, which searches
for the maximum probability of ${\bf X}_t$ given ${\bf y}_{1:t}$.
In a delay-constrained system, one may estimate ${\bf X}_t$ on the
basis of the observations ${\bf y}_{t-\Delta:t+\Delta}$, with
$\Delta$ a fixed interval duration (sliding-window estimator).


%
\subsection{Consideration of a dynamic environment}
Since $\{ {\bf X}_t \}_{t=1}^\infty$ forms a random set sequence,
the statistical characterization of ${\bf X}_t$ is needed for all
discrete time instants $t$. If a dynamic model of the transmission
system is available (which is what we assume in this paper), then
the APPs can be updated recursively, thus allowing one to take
advantage of the information gathered from the past evolution of
the system. We observe in passing that the concept of an adaptive
receiver was examined previously by several authors (see,
e.g.,~\cite{rapbor} and references therein), while the effects on
analysis of a dynamic model were touched upon, among others, by
the authors of~\cite{buikripoo,halbra,mitpoo1,mitpoo2}.

We make the assumption that $\{ {\bf X}_t \}_{t=1}^\infty$ forms a
Markov set sequence, i.e., that ${\bf X}_t$ depends on its past
only through ${\bf X}_{t-1}$. This allows us to use  {\em
Bayesian-filter} recursions for $\widehat{\bf
X}_t$~\cite{mahler2003}:
\begin{eqnarray}
 \lefteqn{ f_{{\bf X}_{t+1}|{\bf  Y}_{1:t}}({\bf B} \mid {  \bf y}_{1:t})} \nonumber \\ &=& \int
 f_{{\bf X}_{t+1}|{\bf X}_t}({\bf  B} \mid
 {\bf  C})f_{{\bf X}_t|{\bf Y}_{1:t}}({\bf  C} \mid {\bf y}_{1:t}) \, \delta {\bf  C}
 \label{rec1}  \\
 \lefteqn{ f_{{\bf X}_{t+1}|{\bf  Y}_{1:t+1}}({\bf  B} \mid {  \bf y}_{1:t+1})} \nonumber \\ &\propto&
 f_{{\bf Y}_{t+1}|{\bf X}_{t+1}}(  {\bf y}_{t+1}
 \mid {\bf B}) f_{{\bf X}_{t+1}|{\bf  Y}_{1:t}}({\bf B}  \mid {\bf y}_{1:t})
 \label{rec2}
 \end{eqnarray}
  The integrals appearing in the equations are {\em set
  integrals}, defined in the sense of RST (see the Appendix).
  The notation $\delta$ for the differential reflects
  this definition.

Thus, the causal maximum-a-posteriori estimate of ${\bf X}_t$ is
obtained by maximizing, over ${\bf B}$, the APP $f_{{\bf X}_t|{\bf
Y}_{1:t}}({\bf B} \mid {\bf y}_{1:t})$, which is tantamount to
minimizing
\[
m({\bf B})\triangleq  ( {\bf y}_t - \sigma({\bf B}))^2  -
\varepsilon({\bf B})
\]
where $\varepsilon({\bf B})\triangleq N_0 \ln f_{{\bf X}_{t} \mid
{\bf Y}_{1:t-1}}( {\bf B} \mid {\bf y}_{1:t-1})$. The first term
in the RHS of definition above is the Euclidean distance between
the observation and the sum of the interfering signals at time
$t$. Its minimization yields ML estimates of ${\bf X}_t$. The
second term in the RHS, generated by the uppermost step of
iterations, reflects the influence on ${\bf X}_t$ of its past
history, and its consideration yields MAP estimates.

 The Bayesian-filter recipe~(\ref{rec1})-(\ref{rec2}) requires two
 ingredients. The first one is the channel model, through the pdf $f_{\bf z}({\bf y}_t\mid {\bf X}_t
 )$. For example, assuming real signals, and the noise to be Gaussian
with mean $0$ and known variance $N_0/2$, we have
\[
f_{\bf z}({\bf y}_t\mid {\bf X}_t )\propto \exp \{ -( {\bf y}_t -
\sigma({\bf X}_t))^2/N_0 \}
\]
The second ingredient is the dynamic model of the random set
sequence ${\bf
 X}_t$, described by the function $f_{{\bf X}_{t+1}|{\bf X}_t}(\, \cdot \, \mid
 \, \cdot \,)$ that describes the time evolution of data and parameters of the system.
 Examples of this modeling procedure
 are available for the problem of tracking multiple targets~\cite{mahler2003,vihola}.

From now on we restrict ourselves to the detection of the number
and identity of active interferers, and of the data they carry,
under the assumption that the remaining parameters, which were
previously estimated by the receiver in a training phase, do not
change in any appreciable way during the tracking phase.
Estimation of these parameters using RST is described in the
companion paper~\cite{angbiglop}.
\section{Detection}\label{detection}
\subsection{Active users}
We assume first that we are only interested in detecting which
interferers, out of a universe of $K$ potential system users, are
present at time $t$. This information may be used for example to
do decorrelation detection, under the assumption that the
signatures of all users are known at the receiver. In our theory,
 ${\bf X}_t$ takes values in $2^{\mathbb K}$. Since this set is finite, a probability measure for
${\bf X}_t$ can be defined by assigning all probabilities
${\mathbb P}({\bf A})$, ${\bf A}\in 2^{\mathbb K}$.

\subsubsection{Static model}
At any fixed time $t$, suppose that the probability of interferer
${\bf x}_t^{(i)}$ to be active is $\alpha$, independent of $t$ and
$i$. In this case the probability of the interferer set ${\bf
X}_t$ depends only on its cardinality $|{\bf X}_t|$, and we can
write
 \beq\label{apriori}
 f_{{\bf X}_t}({\bf  B})=
 \alpha^{|{\bf  B}|}(1- \alpha)^{K- |{\bf  B}|}
 \eeq
To derive this result, we use RST by first computing the belief
function
  \begin{eqnarray}
 \beta_{\bf X}({\bf S}) &\triangleq& {\mathbb P}({\bf X} \subseteq
{\bf S}) \nonumber \\
 &=& \sum_{j=0}^{|{\bf S}|} \sum_{{\bf B}: {\bf B}\subseteq {\bf S} \,\, \& \,\, |{\bf B}|=j} {\mathbb
P}({\bf X}= {\bf B}) \nonumber \\
&=& \sum_{j=0}^{|{\bf S}|} \binom{|{\bf S}|}{j} \alpha^j
(1-\alpha)^{K-j}
 \end{eqnarray}
 and subsequently computing its set derivative, which, in the discrete
 case, becomes the M\"{o}bius inversion
 formula~(\ref{mobius}).

\subsubsection{Dynamic model}
Consider now the evolution of ${\bf X}_t$. We assume that from
$t-1$ to $t$ some new users become active, while some old users
become inactive. We write
 \beq
 {\bf X}_t = {\bf S}_t \cup {\bf N}_t
 \eeq
 where ${\bf S}_t$ is the set of {\em surviving} users still active from $t-1$,
 and ${\bf N}_t$ is the set of {\em new} users becoming active at
 $t$. The condition ${\bf N}_t\cap {\bf
 X}_{t-1}=\emptyset$ is forced, because a user ceasing
 transmission at time $t-1$ cannot re{\"e}nter the set of active users
 at time $t$.
 We proceed by constructing separate dynamic models for
${\bf S}_t$ and ${\bf N}_t$, which will be eventually combined to
yield a model for ${\bf X}_t$.

Consider first ${\bf S}_t$.
 Suppose that there are $k$ active users at $t-1$, the elements of
 the random set ${\bf X}_{t-1}=\{ {\bf x}_{t-1}^{(1)}, \ldots , {\bf
  x}_{t-1}^{(k)}
  \}$. Then we may write, for the set of surviving users,
 \beq\label{surviving_users}
 {\bf S}_t = \bigcup_{i=1}^k {\bf X}_t^{(i)}
 \eeq
 where ${\bf X}_t^{(i)}$ denotes either an empty set (if user $i$ has
 become inactive) or the singleton $\{ {\bf x}_t^{(i)} \}$ (user $i$ is still active). Let
 $\mu$ denote the ``persistence'' probability, i.e., the
 probability that a user survives from $t-1$
 to $t$. We obtain, for the conditional probability of ${\bf S}_t$
 given that ${\bf X}_{t-1}={\bf B} $:
 \beq\label{primo}
 f_{{\bf S}_t|{\bf X}_{t-1}}( {\bf  C} \mid {\bf B}) =
 \left\{ \begin{array}{ll}
\displaystyle
  \mu^{|{\bf  C}|}(1-\mu)^{|{\bf B}|-|{\bf  C}|}, & {\bf
  C}\subseteq {\bf B} \\
  0, & {\bf C}\nsubseteq {\bf B} \end{array} \right.
 \eeq

 For new users, a reasonable model has
 \beq\label{secundo}
f_{{\bf N}_t\mid {\bf X}_{t-1}}( {\bf C} \mid {\bf  B})= \left\{
\begin{array}{ll}
 \alpha^{|{\bf C}|} (1-\alpha)^{K-|{\bf B}|- |{\bf C}|}, & {\bf C} \cap {\bf
 B}=\emptyset \\ 0, & {\bf C} \cap {\bf  B}\neq \emptyset
 \end{array} \right.
 \eeq


 Finally, by assuming that births and deaths of users are conditionally independent given
${\bf X}_{t-1}={\bf B}$, the conditional pdf of the union of the
random sets ${\bf S}_t$ and ${\bf N}_t$ is obtained from the {\em
generalized convolution}~\cite{goodman}
 \begin{eqnarray}
 \lefteqn{ f_{{\bf X}_t|{\bf X}_{t-1}}({\bf C} \mid {\bf B})}
 \nonumber \\
&=&
 \sum_{{\bf  W} \subseteq {\bf C}}
 f_{{\bf S}_t|{\bf X}_{t-1}}({\bf  W} \mid {\bf B}) \, f_{{\bf N}_t|{\bf X}_{t-1}}({\bf C} \setminus
 {\bf W} \mid{\bf B} ) \nonumber \\
 &=& %
f_{{\bf S}_t|{\bf X}_{t-1}}({\bf  C} \cap {\bf  B})f_{{\bf
N}_t|{\bf X}_{t-1}}({\bf  C}\setminus ({\bf  C} \cap {\bf  B}))
 \end{eqnarray}
%

\subsubsection{Bayesian-filter  recursions}
In our context, recursions~(\ref{rec1})-(\ref{rec2}) can be
implemented as follows. Determine first:
\begin{dingautolist}{192}
\item
The a priori probability distribution of ${\bf X}_0$ at the
beginning of the detection process. Description of this
distribution consists of assigning probabilities to all the
elements of $2^{\mathbb K}$. This can be done for example by
assuming independent users with the same stationary activity
factor.
\item
The set of observations ${\bf y}_t$, $t=1, \ldots, T$.
\item
The conditional pdf's $f_{{\bf Y}_t|{\bf X}_t}$, depending on the
channel model.
\item
The ``evolution'' pdf's $f_{{\bf X}_{t+1}|{\bf X}_t}$, depending
on the dynamic model.
\end{dingautolist}
The recursion goes as follows: omitting the subscripts for
notational simplicity here, and identifying random sets with their
realizations, we have
\[
f({\bf X}_1)= \int f({\bf X}_1 \mid {\bf X}_0) f({\bf X}_0) \,
\delta {\bf X}_0
\]
With this, we can compute
\[
f({\bf X}_1 \mid {\bf y}_1) \propto  f({\bf y}_1 \mid {\bf X}_1)
\,  f({\bf X}_1)
\]
which allows the calculation of the causal MAP estimate
$\widehat{\bf X}_1$. Next, we compute
\[
f({\bf X}_2\mid {\bf y}_1)= \int f({\bf X}_2 \mid {\bf X}_1)\,
f({\bf X}_1 \mid {\bf y}_1) \, \delta{\bf X}_1
\]
and hence
\[
f({\bf X}_2 \mid {\bf y}_{1:2})\propto f({\bf y}_2 \mid {\bf X}_2)
\, f({\bf X}_2 \mid {\bf y}_1)
\]
which allows the calculation of $\widehat{\bf X}_2$. The general
recursion has, for $t=2, \ldots$:
\begin{eqnarray*}
f({\bf X}_{t+1} \mid {\bf y}_{1:t}) &=& \int f({\bf X}_{t+1}\mid
{\bf X}_t) \, f({\bf X}_t \mid {\bf y}_{1:t}) \, \delta{\bf X}_t
\\
f({\bf X}_{t+1} \mid {\bf y}_{1:t+1}) &\propto & f({\bf
y}_{t+1}\mid {\bf X}_{t+1}) \, f({\bf X}_{t+1} \mid {\bf y}_{1:t})
\,
\end{eqnarray*}
and, in the case examined in this section,
\begin{eqnarray}
\lefteqn{ f({\bf X}_{t+1} \mid {\bf y}_{1:t})}\nonumber \\ &=&
\sum_{{\bf X}_t \in 2^{\mathbb K}}  f({\bf X}_{t+1}\mid {\bf X}_t)
\, f({\bf X}_t \mid {\bf y}_{1:t}) \label{compl1}
\\
\lefteqn{ f({\bf X}_{t+1} \mid {\bf y}_{1:t+1})}\nonumber \\
&\propto & f_{\bf z}({\bf y}_{t+1}- \sigma({\bf X}_{t+1})) \,
f({\bf X}_{t+1} \mid {\bf y}_{1:t}) \label{compl2}
\end{eqnarray}

\subsection{Active users and their data}
Assume binary information data, independent from time to time and
across users, and a discrete-time unit such that from $t$ to $t+1$
each user transmits $N$ binary symbols. In this case ${\bf X}_t$
takes values in a set with
 \[
 \sum_{k=0}^K \binom{K}{k} 2^{kN} = (1+2^N)^K
 \]
 elements, that we denote $(1+2^N)^{\mathbb K}$.
Eq.~(\ref{apriori}) becomes
 \beq
 f_{{\bf X}_t}({\bf  B})=2^{-N|{\bf B}|}
 \alpha^{|{\bf  B}|}(1- \alpha)^{K- |{\bf  B}|}
 \eeq
 where the new factor $2^{-N|{\bf B}|}$ accounts for the fact that
 there are $N|{\bf B}|$ equally likely binary symbols transmitted
 at time $t$ by $|{\bf B}|$ interferers.

Similarly, (\ref{primo}) is transformed into
 \begin{eqnarray}
 \lefteqn{ f_{{\bf S}_t|{\bf X}_{t-1}}( {\bf  C} \mid {\bf B})}\nonumber \\
 &=&
 \left\{ \begin{array}{ll}
\displaystyle 2^{-N|{\bf C}|}
  \mu^{|{\bf  C}|}(1-\mu)^{|{\bf B}|-|{\bf  C}|}, & {\bf
  C}\subseteq {\bf B} \\
  0, & {\bf C}\nsubseteq {\bf B} \end{array} \right. \label{primo1}
 \end{eqnarray}
and (\ref{secundo}) into
 \begin{eqnarray}
  \lefteqn{ f_{{\bf N}_t\mid {\bf
X}_{t-1}}( {\bf C} \mid {\bf  B})} \nonumber \\&=& \left\{
\begin{array}{ll} 2^{-N|{\bf C}|}
 \alpha^{|{\bf C}|} (1-\alpha)^{K-|{\bf B}|- |{\bf C}|}, & {\bf C} \cap {\bf
 B}=\emptyset \\ 0, & {\bf C} \cap {\bf  B}\neq \emptyset
 \end{array} \right. \label{secundo1}
 \end{eqnarray}

\subsection{Possible scenarios}
We recall that throughout this paper we assume that the only
unknown signal quantities may be the identities of the users and
their data. Specifically, we may distinguish four cases in our
context:
 \begin{dingautolist}{192}
\item
{\em Static channel, unknown identities, known data.} This
corresponds to a training phase intended at identifying users, and
assumes that the user identities do not change during
transmission. In this case we write $\bf X$ in lieu of ${\bf
X}_t$.
\item
{\em Static channel, unknown identities, unknown data.} This may
correspond to a tracking phase following~\ding{192} above. We
write again $\bf X$ in lieu of ${\bf X}_t$, and assume that $\bf
X$ contains the whole transmitted data sequence.
\item
{\em Dynamic channel, unknown identities, known data.} This
corresponds to identification of users preliminary to data
detection (which, for example, may be based on decorrelation).
\item
{\em Dynamic channel, unknown identities, unknown data.} This
corresponds to simultaneous user identification and data detection
in a time-varying environment.
 \end{dingautolist}

\section{An example of application}\label{example}

Assume now the specific situation of a DS-CDMA system with
signature sequences of length $L$ and additive white Gaussian
noise. At discrete time $t$, we may write, for the sufficient
statistics of the received signal,
 \beq\label{cdma}
 {\bf y}_t = {\bf R} {\bf A} {\bf b}_t ({\bf X}_t ) + {\bf z}_t,
 \qquad t=1, \ldots, T
 \eeq
where ${\bf X}_t$ is now the random set of all active users, $\bf
R$ is the $L\times L$ correlation matrix of the signature
sequences (assumed to have unit norm), $\bf A$ is the diagonal
matrix of the users' signal amplitudes, the vector ${\bf b}_t
({\bf X}_t )$ has nonzero entries in the locations corresponding
to the active-user identities described by the components of $
{\bf X}_t  $, and ${\bf z}_t \sim {\EuScript N} (0, (N_0/2){\bf R}
)$ is the noise vector, with $N_0/2$ the power spectral density of
the received noise. We further assume $N=1$, i.e., that at every
discrete time instant only one binary antipodal symbol is
transmitted.

\subsection{Static channel}
The a posteriori probability of $\bf X$, given the whole received
sequence (we omit the time subscript for simplicity), is
 \begin{eqnarray}
 \lefteqn{ f({\bf X} \mid {\bf y}_1, \ldots , {\bf y}_T ) } \nonumber \\
 &\propto&
  f_{\bf X}({\bf X}) \, f({\bf y}_1, \ldots, {\bf y}_T \mid {\bf
  X}) \\  \hspace{-7pt} &=& \hspace{-7pt}
   \exp \left\{ \hspace{-3pt} \displaystyle -\frac{1}{N_0}
  \sum_{t=1}^T \left( {\bf y}_t - {\bf RAb}_t({\bf X})
  \right)^\prime {\bf R}^{-1} \left( {\bf y}_t - {\bf RAb}_t({\bf
  X}) \right) \hspace{-3pt} \right\} \nonumber \\
  & & \times f_{\bf X}({\bf X}) \nonumber
 \end{eqnarray}
 Thus, the MAP estimator of users' identities is
 \beq
 \widehat{\bf X} = \arg \max_{{\bf X} \in 2^{\mathbb K}}
 f({\bf X} \mid  {\bf y}_{1:T} )
 \eeq
 where, as usual, ${\bf y}_{1:T} \triangleq {\bf y}_1, \ldots , {\bf y}_T$.
 The MAP estimator of users' identities and data is
 \beq
 \widehat{\bf X} = \arg \max_{\bf X}
 f({\bf X} \mid {\bf y}_{1:T} )
 \eeq
 {
 where the set of possible realizations of $\bf X$ includes
 $(1+2^T)^K$ elements: in fact, in $T$ time interval the number
 of transmitted binary symbols is $2^{|{\bf X}|T}$, and
 \[
 \sum_{|{\bf X}|=0}^K \binom{K}{|{\bf X}|}2^{|{\bf X}|T}=(1+2^T)^K
 \]
}
 The expression above can be rewritten in such a way that the
presence of the sequence of transmitted data is made more
explicit. Specifically, we may write, in lieu of $\bf X$,
 the sequence $( {\bf X}, {\bf b}_1({\bf X}), \ldots,
 {\bf b}_T({\bf X}))$. Doing so, we may express the MAP
 estimator of users' identities and data in the more explicit form
 \begin{eqnarray}
\lefteqn{ (\widehat{\bf X}, \widehat{\bf b}_1({\bf X}), \ldots,
\widehat{\bf b}_T({\bf
 X}))} \nonumber \\
 &=& \arg \max f({\bf X}, {\bf b}_1({\bf X}), \ldots, {\bf b}_T({\bf X})\mid {\bf y}_{1:T})
 \end{eqnarray}
 where the maximum has to be taken with respect to ${\bf X}$
 and ${\bf b}_t ({\bf X})$, $t=1, \ldots, t$. The introduction of this
 ``fine-grain'' notation for the random set suggests that the MAP detector may be
 implemented in the form of a sequential detector, thus simplifying its operation
 (more on this {\em infra}).

 {\color{black} Before moving on,
 it may be worth pointing out that, for the case considered here,
 the same results could be obtained through ordinary probability
 theory by introducing a $K$-dimensional vector sequence,
 taking on binary or ternary values according to whether a
 training sequence is used or not: as already anticipated, though,
 such equivalence holds only for the discrete case, and, unlike the
 RST formulation, does not suggest an immediate extension
 accounting for more general channel models~\cite{angbiglop}.}

\subsection{Dynamic channel}
 Consider now a dynamic channel, and examine first the case of
 known data. We have, accounting for the Markov property of our
 channel model,
 \begin{eqnarray}
 \lefteqn{ f({\bf X}_1, \ldots, {\bf X}_T \mid {\bf y}_{1:T})} \\
 &\propto&
 f({\bf y}_{1:T} \mid {\bf X}_1, \ldots , {\bf X}_T)\times
f({\bf X}_1) \prod
 _{t=2}^T f({\bf X}_t \mid {\bf X}_{t-1}) \nonumber
 \end{eqnarray}
 with $f({\bf X}_1)$ a density whose assignment is based upon prior
 knowledge of the channel state at the beginning of the
 transmission.
 The MAP estimator here maximizes the RHS of the above (or its
 logarithm) with respect to the values taken on by the
 sequence $({\bf X}_1, \ldots, {\bf X}_T )$.
 Even in this case we may think of a sequential detector, which searches
 for the maximum-APP path traversing a
 trellis having $T$ stages and a number of states at stage $i$ equal to the number of
 realizations of the random set ${\bf X}_i$.

\paragraph{Implementing a sequential detector.}
 Implementation of
the sequential detector through a version of Viterbi algorithm
leads to the following consequences:
\begin{dingautolist}{192}
\item
The decision on the whole sequence of users' identities and their
data should be taken only after the whole sequence of observations
${\bf y}_1, \ldots , {\bf y}_T$ has been recorded.
\item
The decision on the users' identities and their data at time $t$
depends not only on the past observations, but also on
observations that have not been recorded yet at time $t$.
\item
A suboptimum version of the optimum sequential algorithm, the {\em
 sliding-window Viterbi} algorithm (see, e.g., ~\cite[p.\ 133 ff.]{codfad})
 can be implemented. This consists of forcing a decision on ${\bf X}_t$, ${\bf b}_t({\bf X}_t)$
 based on a sliding window of observations that includes ${\bf
 y}_t$, but whose length is smaller than $T$.
\end{dingautolist}

\subsection{PEP analysis}
We now evaluate the performance of the detectors described above.
We assume $N=1$ for simplicity, and derive bounds and
approximations to error probabilities using the pairwise error
probability (PEP) $P({\bf X}_t \rightarrow \widehat{\bf X}_t)$.
This is the probability that, when ${\bf X}_t$ is the true value
of the random set to be detected, the receiver assigns a higher
APP to $\widehat{\bf X}_t \neq {\bf X}_t$ (see, e.g.,~\cite[p.\
43] {codfad})~\footnote{It might be worth observing here that,
contrary to a fairly widespread misconception, $P({\bf X}_t
\rightarrow \widehat{\bf X}_t)$ is {\em not} the probability of
mistaking $\widehat{\bf X}_t$ for ${\bf X}_t$, unless ${\bf X}_t$
and $\widehat{\bf X}_t$ are the only possible alternatives.}

\subsubsection{Static channel}
Defining
 \begin{eqnarray}\label{eq30}
 \lefteqn{ {\bf S}_t ({\bf X}, \widehat{\bf X} )  } \\ \hspace{-32pt} &\triangleq&
 \hspace{-8pt}
 {\bf R}^{-1} \hspace{-4pt} \left[  \bigl( {\bf y}_t- {\bf RAb}_t(\widehat{\bf X})
 \bigr) \bigl({\bf y}_t- {\bf RAb}_t(\widehat{\bf X})
 \bigr)^\prime \right] \biggr|_{{\bf y}_t = {\bf RAb}_t ({\bf
 X})+{\bf z}_t} \nonumber
 \end{eqnarray}
 we have
 \beq
{\bf S}_t ({\bf X}, {\bf X} )= {\bf R}^{-1}{\bf z}_t {\bf
z}_t^\prime
 \eeq
 and
 \beq
{\bf S}_t ({\bf X}, \widehat{\bf X} )= {\bf R}^{-1} \bigl[ \bigl(
{\bf RAd}_t({\bf X}, \widehat{\bf X})+{\bf z}_t \bigr) \bigl( {\bf
RAd}_t({\bf X}, \widehat{\bf X})+{\bf z}_t \bigr)^\prime \bigr]
 \eeq
 where ${\bf d}_t({\bf X}, \widehat{\bf X})\triangleq {\bf
 b}_t({\bf X})-{\bf b}_t(\widehat{\bf X})$.
Based on the above, the PEP with ML detection of unknown user
identities can be written as
 \beq
P({\bf X}  \rightarrow \widehat{\bf X} )= {\mathbb P} \left\{
\mbox{tr}\, \left[ \sum_{t=1}^T {\bf S}_t ({\bf X} , \widehat{\bf
X}  )- {\bf S}_t ({\bf X} , {\bf X}  ) \right] <0 \right\}
 \eeq
Now, observe that
 \begin{eqnarray*}
 \lefteqn{ \mbox{tr}\, \left[{\bf S}_t ({\bf X} , \widehat{\bf
X}  )- {\bf S}_t ({\bf X} , {\bf X}  )\right]} \nonumber \\ &=&
\mbox{tr}\, \left[ {\bf R}^{-1} \left( {\bf RAd}_t{\bf d}_t^\prime
{\bf AR} + {\bf z}_t{\bf d}_t^\prime  {\bf AR}+{\bf RAd}_t{\bf
z}_t^\prime \right)\right] \\
 &=& \mbox{tr}\,\left[ {\bf Ad}_t{\bf d}_t^\prime{\bf AR}+{\bf
R}^{-1}{\bf z}_t{\bf d}^\prime_t{\bf AR}+{\bf Ad}_t{\bf
z}_t^\prime\right] \\
&=& \mbox{tr}\, \left[ {\bf Ad}_t{\bf d}^\prime_t {\bf AR}+{\bf
z}_t{\bf d}_t^\prime {\bf A} + {\bf Ad}_t {\bf z}_t^\prime \right]
\\
&=& \mbox{tr}\,\left[ {\bf RAd}_t{\bf d}_t^\prime {\bf A}\right] +
2 \, \mbox{tr}\, \left[ {\bf Ad}_t{\bf z}^\prime_t \right]
 \end{eqnarray*}
 Denoting by $a(i)$ the $i$th
 diagonal element of matrix $\bf A$, by $d_t(i)$ the $i$th entry of ${\bf d}_t$, by
 $r_{j,k}$ the entry in row $j$ and column $k$ of $\bf R$, and by
 $z_t(i)$ the $i$th entry of vector ${\bf z}_t$, we have
 \begin{eqnarray}
 \mbox{tr}\,\left[ {\bf RAd}_t {\bf d}_t^\prime {\bf A}\right] &=&
 \sum_{i=1}^K\sum_{j=1}^K a(i) a(j) d_t(i) d_t(j) r_{i,j} \\
\mbox{tr}\,\left[ {\bf Ad}_t {\bf z}_t^\prime \right] &=&
\sum_{i=1}^K a(i) d_t(i) z_t(i)
 \end{eqnarray}
 Finally, since we are assuming ${\bf z}_t \sim {\EuScript N}({\bf
 0}, (N_0/2){\bf R})$, we have
 \beq
 \mbox{tr}\, \left[\sum_{t=1}^{T}\left({\bf S}_t ({\bf X} , \widehat{\bf
X}  )- {\bf S}_t ({\bf X} , {\bf X}  )\right)\right]\sim
{\EuScript N}(\xi_T, 2N_0 \xi_T)
 \eeq
where
 \beq
 \xi_T \triangleq \sum_{t=1}^T \sum_{i=1}^K\sum_{j=1}^K
 a(i) a(j) d_t(i) d_t(j) r_{i,j}
 \eeq
 In conclusion, we obtain
 \beq
 P({\bf X} \rightarrow \widehat{\bf X})
= Q \left( {\sqrt
 \frac{\xi_T}{2N_0}} \right)
 \eeq
 where $Q( \, \cdot \, )$ denotes the Gaussian tail function.

Before proceeding further, we comment briefly on the structure of
vectors ${\bf d}_t$. They have the following nonzero entries:
 \begin{dingautolist}{192}
 \item
 The  $|{\bf X} \cap \widehat{\bf X}|$ terms corresponding to
 users present in both sets: these terms may take on values in $\{
 0, \pm 2 \}$.
 \item
 The $| \widehat{\bf X} \setminus {\bf X} \cap \widehat{\bf
 X} |$ terms corresponding to users present in $\widehat{\bf
 X}$ only: these terms may take on values in $\{ \pm 1 \}$.
 \item
 The $| {\bf X} \setminus {\bf X}\cap \widehat{\bf
 X} |$ terms corresponding to users present in ${\bf
 X}$ only: these terms may take on values in $\{ \pm 1 \}$.
 \end{dingautolist}

 Similarly, the PEP with MAP detection has the form
 \beq
 P({\bf X} \rightarrow \widehat{\bf X})  =
Q \left(
 \frac{\xi_T-\eta}{\sqrt{{2N_0 \xi_T}} } \right) \label{aux}
 \eeq
where
 \[
 \eta \triangleq N_0 \ln \left[ \frac{f_{{\bf X}}(\widehat{\bf X})}{f_{{\bf X}}({\bf X})} \right]
 \]
 Observe here that, with $\bf R$ a positive definite matrix,
 we have
 \[ ({\bf Ad}_t)^\prime {\bf R} ({\bf Ad}_t) >0 \quad \mbox{for } {\bf
 Ad}_t \neq {\bf 0}
 \]
 which entails
 \[
 \lim_{T\rightarrow \infty}P({\bf X} \rightarrow \widehat{\bf
 X})=0
 \]
 Notice also that, for given signal-to-noise ratio, (\ref{aux})
 also suggests a minimum length of the data frame in the form of the ``open-eye''
 condition $T \geq T_{\rm min}$ , where
 \[
 T_{\rm min} \triangleq \inf \left\{ T: \min_{{\bf X},\widehat{\bf
 X}} \bigl[ \xi_T({\bf X},\widehat{\bf X})-\eta({\bf X},\widehat{\bf
 X}) \bigr] >0 \right\}
 \]

 The PEP for the case of detection of user identities and data can be
 dealt with with similar techniques, and we shall not delve in
 this issue any further here.

\subsubsection{Dynamic channel}
In this case, defining the true state sequence $\underline{\bf
X}\triangleq ({\bf X}_t)_{t=1}^T$ and the competing state sequence
$\underline{\widehat{\bf X}} \triangleq (\widehat{\bf
X}_t)_{t=1}^T$, we obtain the PEP for the MAP detection of unknown
identities: \beq
 P(\underline{\bf X}\rightarrow \underline{\widehat{\bf X}})= Q
 \left(
      \frac
      {\xi_T (\underline{\bf X},\underline{\widehat{\bf  X}})-\eta_T(\underline{\bf X},\underline{\widehat{\bf X}})}
      {\sqrt{2N_0\xi_T (\underline{\bf X},\underline{\widehat{\bf X}})}}
 \right)
 \eeq
 and the PEP for the MAP detection of unknown identities and
data:
 \begin{eqnarray}
 \lefteqn{ P\bigl(\underline{\bf X}, ( {\bf b}_t ( {\bf X}_t))_{t=1}^T \rightarrow \underline{\widehat{\bf
 X}}, ({\bf b}_t(\widehat{\bf X}_t))_{t=1}^T\bigr)} \nonumber \\ &=&  Q
 \left(
      \frac
      {\xi_T (\underline{\bf X},\underline{\widehat{\bf  X}})-\tilde{\eta}_T(\underline{\bf X},\underline{\widehat{\bf X}})}
      {\sqrt{2N_0\xi_T (\underline{\bf X},\underline{\widehat{\bf X}})}}
 \right)
 \end{eqnarray}
where
\begin{eqnarray}
\xi_T (\underline{\bf X},\underline{\widehat{\bf  X}})
&\triangleq& \sum_{t=1}^T {\bf d}^\prime_t {\bf ARAd}_t \\
\eta_T(\underline{\bf X},\underline{\widehat{\bf X}}) &\triangleq&
N_0 \sum_{t=1}^T \ln \left[ \frac {f_{{\bf X}_t|{\bf
X}_{t-1}}(\widehat{\bf X}_t \mid \widehat{\bf X}_{t-1})} {f_{{\bf
X}_t|{\bf X}_{t-1}}({\bf X}_t \mid {\bf X}_{t-1})} \right] \\
\tilde{\eta}_T(\underline{\bf X},\underline{\widehat{\bf X}})
&\triangleq& \eta_T(\underline{\bf X},\underline{\widehat{\bf X}})
\nonumber \\ & &+N_0 \sum_{t=1}^T  \left[ |{\bf X}_t| -
|\widehat{\bf X}_t| \right] \ln 2
\end{eqnarray}

Similar arguments, which we omit here for brevity's sake, apply to
ML detection.

\subsection{Error probabilities}
Several approximations to error probabilities are possible, based
on the union bound (see, e.g.,~\cite{codfad}), on the PEP
derivations outlined {\em supra}, and on assumptions on user
statistics, spreading codes, and users' amplitudes. We obtain, for
the union bound to the probability of mistaking the set of active
users,
 \beq
 P(e) \leq \sum_{i=1}^{2^K} f({\bf X}_i) \sum_{j\neq i}
P( {\bf X}_i \rightarrow {\bf X}_j)
 \eeq
which, under maximum prior uncertainty as to the channel
occupancy, becomes:
  \beq
 P(e) \leq \frac{1}{2^K} \sum_{i=1}^{2^K}
\sum_{j\neq i} P( {\bf X}_i \rightarrow {\bf X}_j)
 \eeq
 where ${\bf X}_i, {\bf X}_j \in 2^{{\mathbb K}}$. This union bound
can be simplified by restricting the inner summation to those
pairs of realizations of the random sets that are most likely to
contribute significantly to error probability. For example, if we
restrict it to the pairs that differ in at most $n$ entries, we
obtain an approximation depending on $n$:
 \beq\label{approxx}
P(e) \approx  P^{(n)}(e) \triangleq \frac{1}{2^K} \sum_{i=1}^{2^K}
 \sum_{j:\delta_{i,j}\leq n} P( {\bf X}_i \rightarrow {\bf X}_j)
 \eeq
where
 \[
 \delta_{i,j} \triangleq \left| {\bf X}_i \setminus {\bf X}_i \cap {\bf
 X}_j \right| + \left| {\bf X}_j \setminus {\bf X}_i \cap {\bf
 X}_j \right| \leq n
 \]
Likewise, for a dynamic scenario, we have that the union bound for
user identification is written as: \beq P(e) \leq
\sum_{\underline{\bf X}\in (2^{{\mathbb K}})^T}f(\underline{\bf
X})\sum_{\underline{\bf X}\neq \underline{\widehat{\bf X}}}
P(\underline{\bf X}\rightarrow \underline{\widehat{\bf X}})\eeq
where $f(\underline{\bf X})$ can be easily determined by applying
the chain rule to the Markov set sequence $\underline{\bf X}$.
Approximations similar to (\ref{approxx}) can be developed;
likewise, the case of joint user identification and data detection
can be handled by noticing that the configurations of
$\underline{\bf X}$ become now $3^{KT}$, and the joint density
$f(\underline{\bf X})$ is written as: \beq f(\underline{\bf
X})=2^{-|{\bf X}_1|}f({\bf X}_1)\prod_{t=2}^{T} 2^{-|{\bf X}_t|}
f({\bf X}_t \mid {\bf X}_{t-1}) \label{Markov}\eeq

The above relationships also suggest a semi-analytical method to
evaluate the approximations without summing up an exponential
number of terms: indeed, since an average over the joint density
(\ref{Markov}) is to be performed, this can be efficiently
evaluated through Monte-Carlo counting by generating a
substantially smaller number of independent set sequence patterns
obeying the Markov law (\ref{Markov}).

\section{Numerical results}\label{results}

In this section we show some numerical examples that illustrate
the theory developed before.

Fig.~\ref{new_figure_3} shows how the knowledge of the channel
dynamics can improve the performance of a multiuser detector.
\insertfig{t}{0.8}{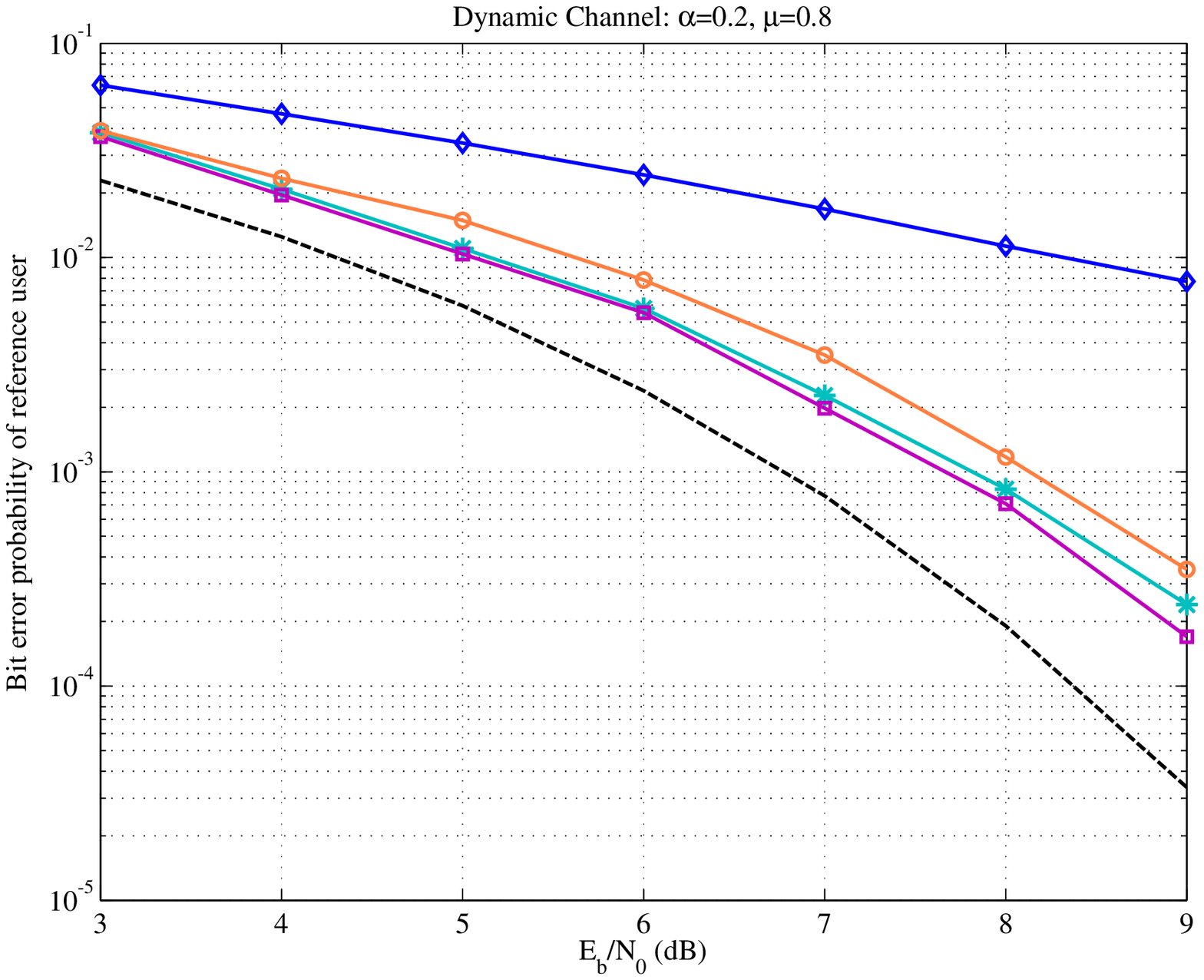}{\sl Bit error probability of the
reference user in a multiuser system with $2$ interferers,
following a dynamic model described above with $\alpha=2$ and $\mu
=0.8$. Line with diamond markers: Classic multiuser ML detection,
assuming that all users are active. Line with circle markers: MAP
detection based on the knowledge of $\alpha$ alone. Line with star
markers: causal RST detector, based on Bayes recursions. Line with
square markers: Viterbi RST detector. Dashed curve: Single-user
bound.}{new_figure_3}

Fig.~\ref{Trained_static_n_6_p_3} refers again to a static channel
and to the case that the active users transmit a known sequence of
bits in order to be identified: we assume here that all users
(including the reference user) are active with probability $\alpha
=0.5$. Now $K=6$, the transmitted signals are binary antipodal,
spreading is done through $m$-sequences with length $7$, the power
control is perfect (hence, $\bf A$ is a scalar matrix) and the
data-frame length varies from $T=1$ to $T=3$. Here we evaluate the
accuracy of the union bound to the probability of an error in the
identification of active users (set-error probability, or SEP),
and of its approximation $P^{(1)}(e)$ (obtained by assuming that
the errors can only be generated by the event, denoted ${\EuScript
E}(1)$, of mistaking an active-user set by another differing by
only one of its elements). Simulation results are also shown for
reference's sake. It can be observed how, especially for large
values of signal-to-noise ratio, the error probability is
dominated by the event ${\EuScript E}(1)$.
\insertfig{h}{0.8}{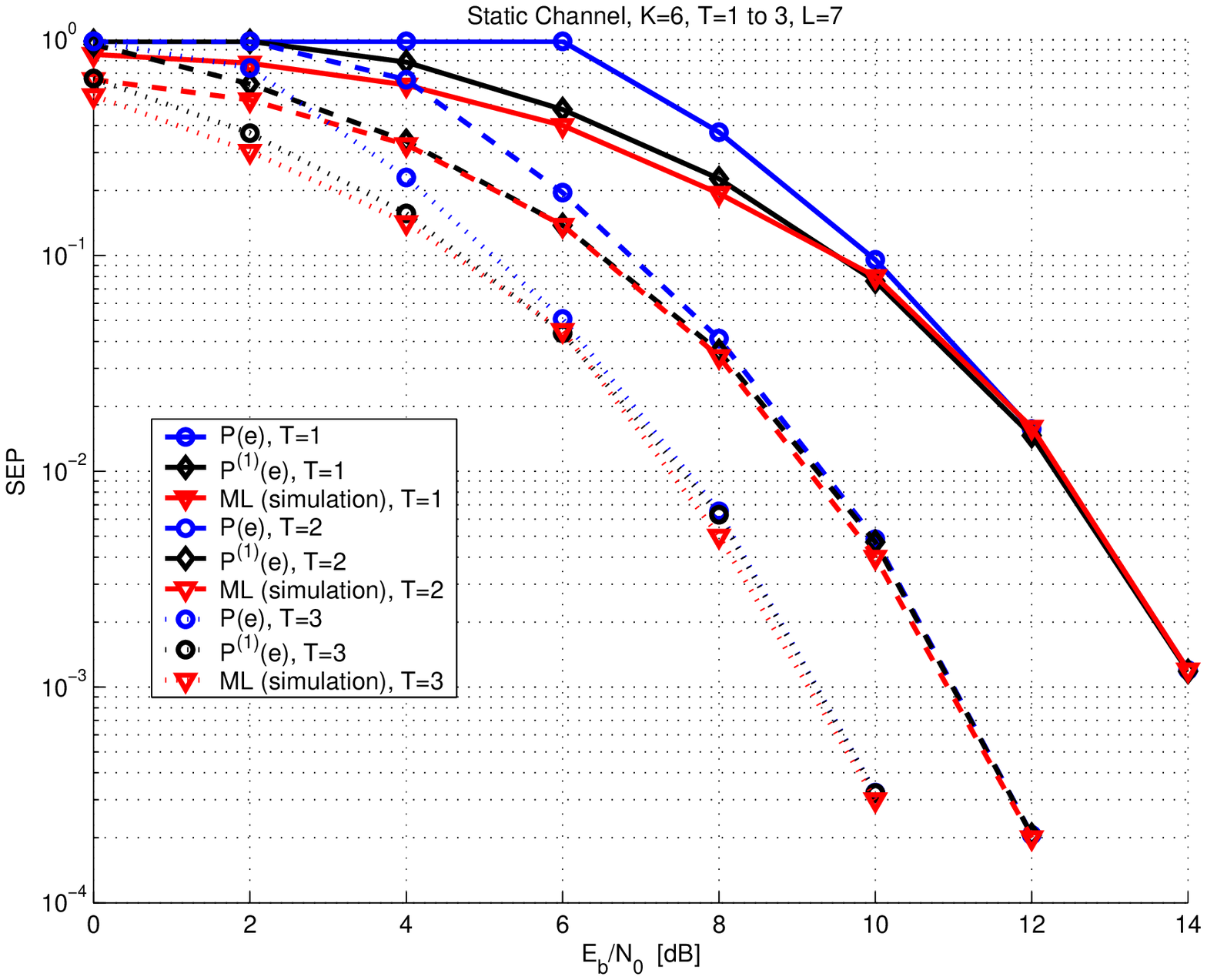}{\sl Set-error
probability with ML detection based on RST with $K=6$, $L=7$.
Comparison among ``exact'' probability (obtained by simulation),
union-bound to it (denoted $P(e)$), and
approximation~(\ref{approxx}) to the union
bound.}{Trained_static_n_6_p_3}

Fig~\ref{Viterbi_vs_Bayes} refers to a system with the same
configuration as in Fig.~\ref{Trained_static_n_6_p_3}, but on a
dynamic channel with $K=3$, $\alpha=0.2$ and $\mu=0.8$. The system
dynamics are tracked over an interval with length $T=10$. The
ordinate shows the set error probability (SEP), i.e., the
probability of an erroneous estimate of the active-user sequence.
Here a comparison is made between a non-causal Viterbi set
estimate and a causal estimate, obtained through Bayesian-filter
recursions. To elicit the impact of the causality constraint, we
represent the SEP for the set ${\bf X}_1$, where the causality
constraint prevents sequence detection, and for the set ${\bf
X}_{10}$, where such a constraint has no effect: as expected, the
performances of the Viterbi algorithm and of the Bayesian
recursions coincide when estimating ${\bf X}_{10}$, while the
causality constraint has a perceivable effect on the performance
when ${\bf X}_{1}$ is estimated.
\insertfig{h}{0.8}{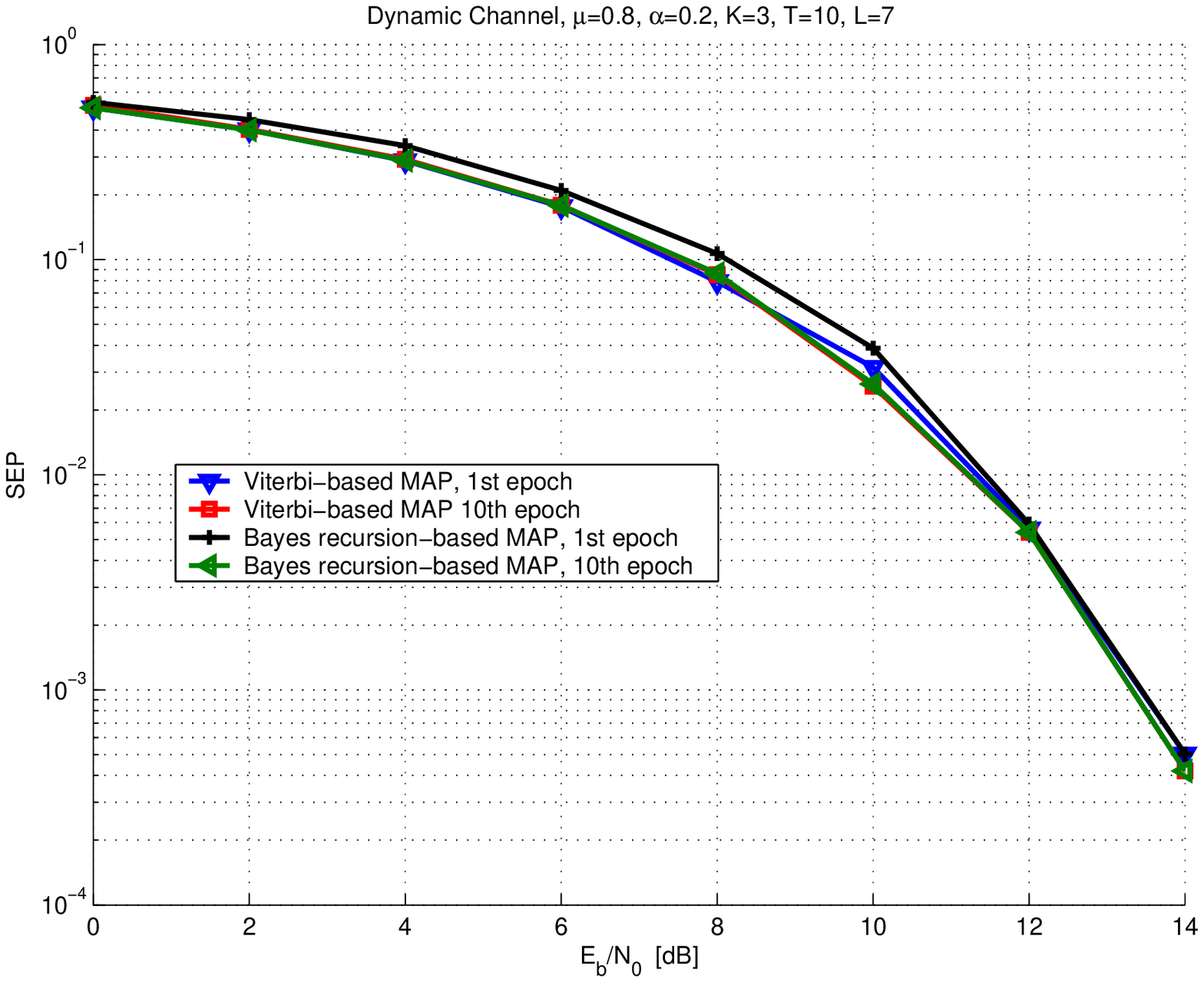}{\sl Trained acquisition of
the set of active users through the Viterbi Algorithm and Bayes
recursion: effect of the causality constraint}{Viterbi_vs_Bayes}

In order to provide global figures of merit of both trained and
untrained systems in a dynamic environment, we use the ``Set
Sequence Error Probability''(SSEP). For trained systems, this is
the probability that for some $t$, $1 \leq t \leq T$, the
estimated set $\widehat{{\bf X}}_t$ differs from the true set
${\bf X}_t$ either in its cardinality or in its elements. For
untrained systems,  it is the probability that at some $t$ the
estimated and the true set differ either in the cardinality and/or
in the identities of the active users and/or in the transmitted
data. Plots of the SSEP are shown in
Fig.~\ref{Trained_dynamic_n_6_p_3} for a trained system with $K=6$
maximum number of active users: also shown in the figure is the
curve obtained through the semi-analytical approximation suggested
in the previous section, which apparently follows the numerical
results quite closely.
\insertfig{h}{0.8}{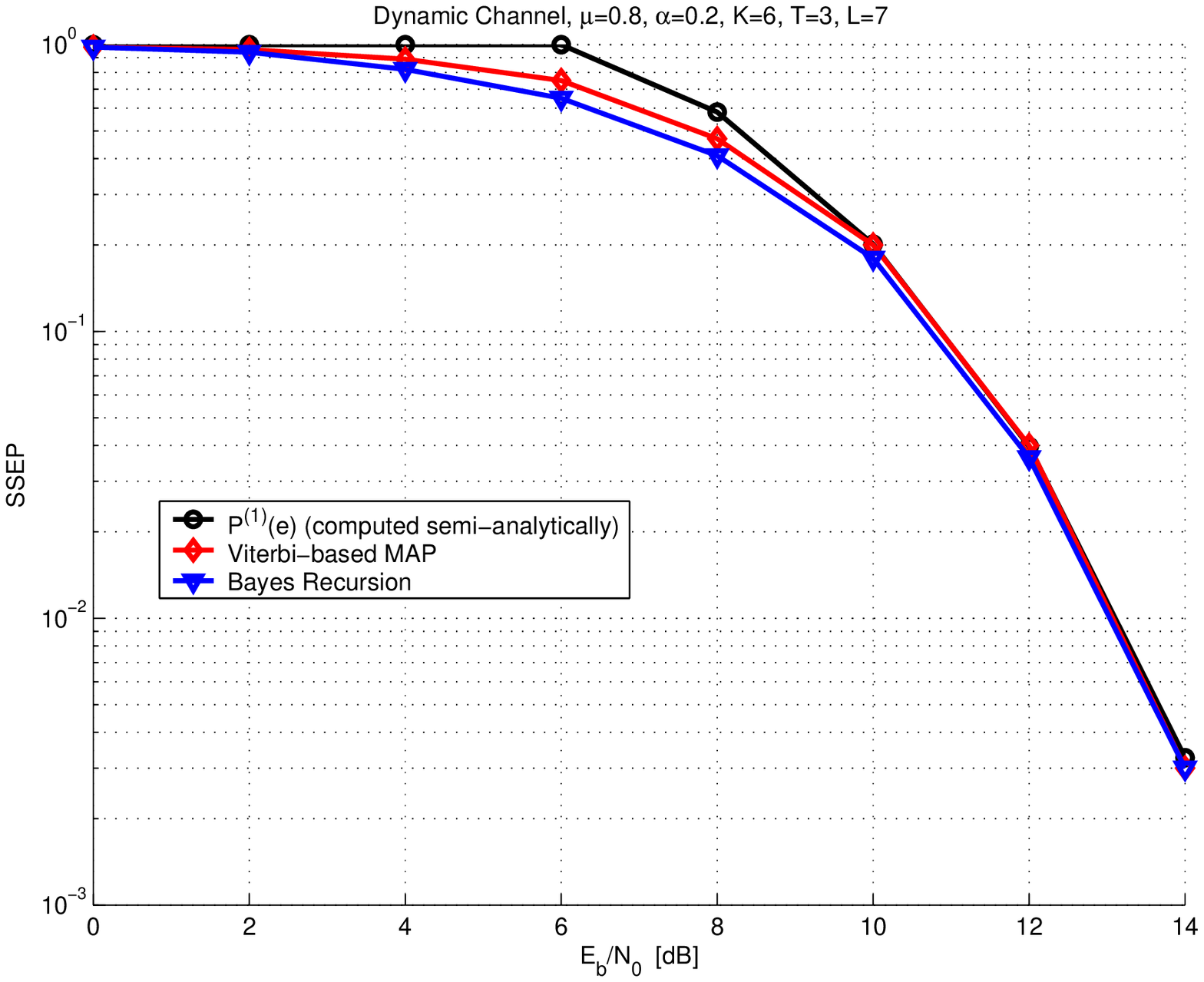}{\sl Set sequence error
probability with $K=6$, $L=7$. Also, the curve corresponding to a
semi-analytical performance evaluation under trained acquisition
of the set of the active users.}{Trained_dynamic_n_6_p_3}

The case that not only the identities, but also the data of the
active users are to be estimated is shown in
Fig.~\ref{Viterbi_vs_bayes_blind1}, assuming a maximum of $K=3$
active users and, again, $L=7$; the data-frame length is $T=10$.
Here we compare a Viterbi-algorithm receiver with one based on
Bayesian recursions for estimating the set of interferers and the
transmitted bits. The ordinate shows the bit-sequence error
probability, at time $t=1$ and at time $t=T=10$, defined as the
probability that the estimated and the true set do not coincide:
the term "bit sequence error probability" is tied here to the fact
that an error in estimating the identities of the active users
automatically implies an error in estimating the stream ${\bf
b}({\bf X}_t)$, while the converse is not true. Once again, the
effect of the causality constraint on the performance is elicited,
and the results are in accordance with the intuition as well as
with the curves of Fig.~\ref{Viterbi_vs_Bayes}.
\insertfig{h}{0.8}{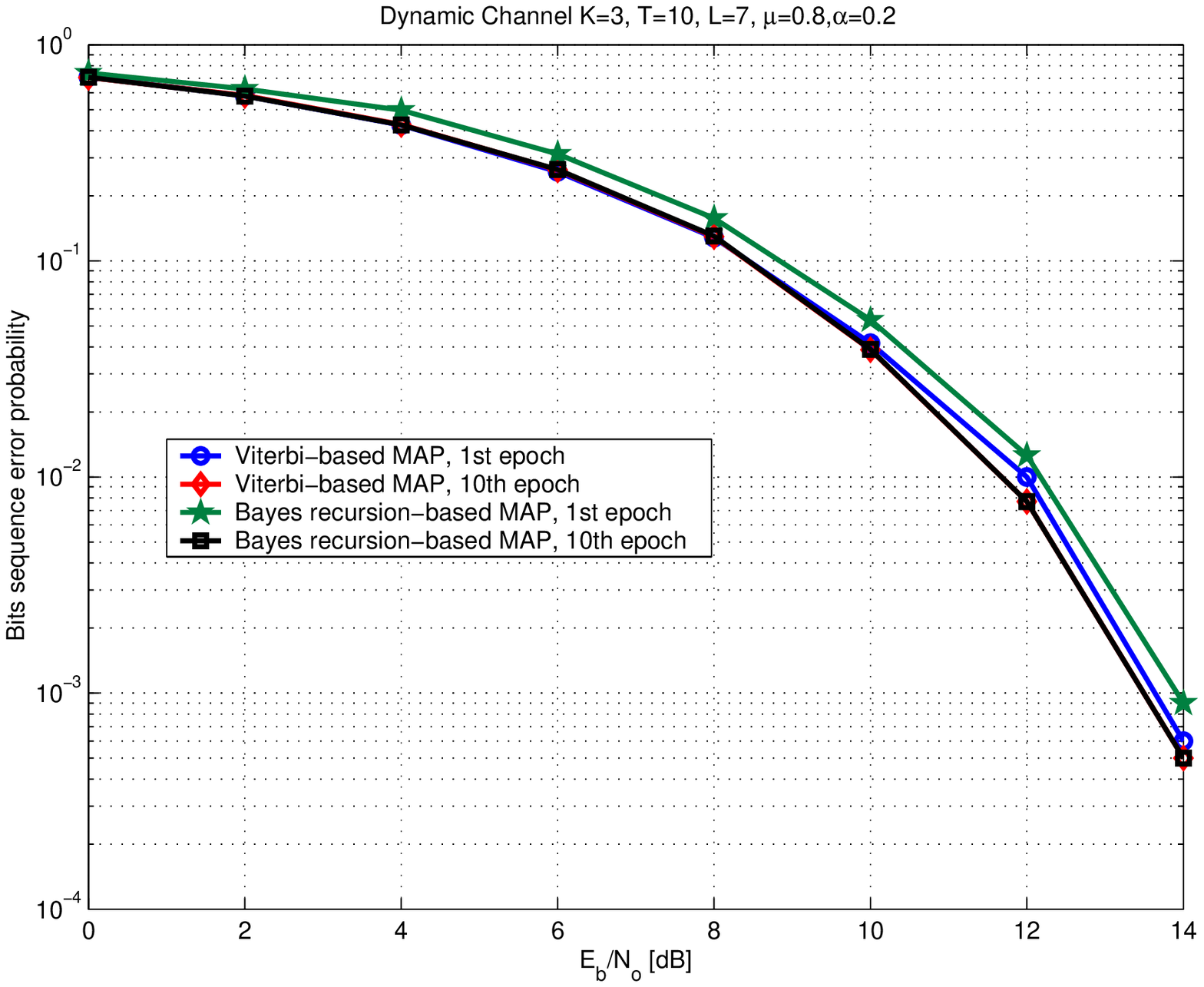}{\sl Bit-sequence error
probability of the reference user. Data estimated with Viterbi
algorithm and Bayesian recursions.}{Viterbi_vs_bayes_blind1}

\section{Conclusions}

We have described a technique for estimating the received-signal
parameters and data in a multiuser transmission system. Since the
number of active interferers is itself a random variable, the set
of parameters to be estimated has a random number of random
elements. A dynamic model for the evolution of this random set
accounts for new interferers appearing and old interferers
disappearing in each measurement interval. {\color{black}
Random-set theory can be used to develop a multiuser detection
scheme in this context. This is done by developing
Bayesian-filtering equations that describe the evolution of the
MAP multiuser detector in a dynamic environment.}

{
One question that may naturally arise at this point
concerns the need for RST, and in particular the question if the
results obtained with RST can also be obtained in a more
straightforward way. Although other ways of obtaining the same
results with conventional probability techniques may be available
(see,e.g.,~\cite{vosindou}, where the connections between RST and
point processes are explored), we argue that not only the rigor
and the generality of RST, but especially its simplicity, make it
a tool of choice for the study of random-access systems. The
results of this paper are also meant to support this claim. }


\appendix

\section{Random-set theory}

This appendix describes, in a rather qualitative fashion, the
fundamentals of Random-Set Theory. For a rigorous approach and
additional details, see~\cite{goodman,vihola,vosindou} and the
references therein.

 Given a sample space $\Omega$ (the space of all the
outcomes of a random experiment), a probability measure can be
defined on it. If a random variable (i.e., a mapping from $\Omega$
to another space $\mathbb S$) is defined, it is convenient to
generate a probability measure directly on $\mathbb S$. This can
be given in terms of a density function, once certain mathematical
operations, such as integration, are defined on $\mathbb S$.
Random sets can be viewed as a generalization of the concept of a
random variable. A {\em finite random set} is a mapping ${\bf X}:
\Omega \rightarrow {\EuScript F}({\mathbb S})$ from the sample
space $\Omega$ to the collection of closed sets of the space
$\mathbb S$, with $|{\bf X}(\omega)|<\infty$ for all $\omega \in
\Omega$. For our purposes, the space $\mathbb S$ of finite random
sets is assumed to be the {\em hybrid space} ${\mathbb S}={\mathbb
R}^d \times U$, the direct product of the $d$-dimensional
Euclidean space ${\mathbb R}^d$ and a finite discrete space $U$.
The elements of $\mathbb S$ characterize the users' parameters,
some of which continuous ($d$ real numbers) and some discrete (for
example, the users' signatures and their transmitted data). An
element of $\mathbb S$ is the pair $({\bf v},u)$, $\bf v$ a
$d$-dimensional real vector, and $u\in U$.


\subsection{Belief mass functions}
A fairly natural probability law for $\bf X$ is the probability
distribution $P_{\bf X}$, defined for any (Borel) subset
$\EuScript T$ of ${\EuScript F}({\mathbb S})$ by
\[
 P_{\bf X}({\EuScript T}) \triangleq {\mathbb P}(  {\bf
 X} \in {\EuScript T})
\]
However, RST is based on a probability law given differently.
Specifically, the {\em belief mass function} of a finite random
set $\bf X$ is defined as
 \beq
 \beta_{\bf X}({\bf C}) \triangleq {\mathbb P}({\bf X} \subseteq {\bf C})
 \eeq
where ${\bf C}$ is a closed subset of $\mathbb S$. As observed
in~\cite[p.\ 42]{vihola}, the belief function corresponds to the
cumulative distribution function of a real random variable, and
differs from it because subsets are only partially ordered by the
inclusion relation $\subseteq$. The belief function characterizes
the probability distribution of a random finite set $\bf X$, and
allows the construction of a density function of $\bf X$ through
the definition of a {\em set integral} and a {\em set derivative}.
Specifically, the {\em belief density}, i.e., the {\em set
derivative} of the belief function, plays the role of a
probability density function in ordinary probability calculus (for
this reason, in this paper we refer to it simply as a {\em
density}). The belief density is obtained as
 \beq\label{setder}
 f_{\bf X}( {\bf Z}) = \left. \frac{\delta \beta ({\bf S})}{\delta {\bf
Z}} \right|_{{\bf S}=\emptyset}
 \eeq
 where $\delta$ denotes the set derivative, to be defined below.
 As observed in~\cite[p.\ 163]{goodman}, the value $\beta_{\bf X}({\bf
Z})$ of the belief density specifies the likelihood with which the
random set $\bf X$ takes the set $\bf Z$ as a specific
realization.

Notice how, in the special case of a random set consisting of a
singleton, ${\bf X} = \{ {\bf x} \}$, $\bf x$ a random vector, we
have
 \[
\beta_{\bf X}({\bf C}) \triangleq {\mathbb P}({\bf X} \subseteq
{\bf C}) = {\mathbb P} ( {\bf x} \in {\bf C}) = P({\bf C})
 \]
 with $P (\, \cdot \,)$ the ordinary probability measure of $\bf x$.

\subsection{Set derivative} Let ${\EuScript C}({\mathbb S})$
denote the collection of closed subsets of $\mathbb S$. If $F$ is
a set function defined on ${\EuScript C}({\mathbb S})$, then its
set derivative at ${\bf x}$ is defined as the set function
\[
\frac{\delta { F} ({\bf S})}{\delta {\bf x}}= \lim_{j \rightarrow
\infty} \lim_{i \rightarrow \infty} \frac{F\left[ \left( {\bf S}
\setminus B_x(1/j)\right) \bigcup
\overline{B_x}(1/i)\right]-F({\bf S})}{\overline{m}\left(
\overline{B_x}(1/i)\right)}
\]
where $B_x(1/j)$, $\overline{B_x}(1/i)$ are an open ball of radius
$1/j$ and a closed ball of radius $1/i$, respectively, both
centered at $\bf x$, and $\overline{m}(\cdot )$ denotes the hybrid
Lebesgue measure, i.e.,
 the product of the ordinary measure in ${\mathbb R}^d$
 and of the counting measure.
 The set derivative of $F$ at a finite set ${\bf X} = \{ {\bf x}_1,
 \ldots, {\bf x}_n \}$ is defined by the recursion
 \[
 \frac{\delta F({\bf S})}{\delta {\bf X}} \triangleq \frac{\delta}{\delta
 {\bf x}_n} \left( \frac{\delta F({\bf S})}{\delta  \{ {\bf x}_1, \ldots ,
 {\bf x}_{n-1} \} }\right)
 \]
In particular, the belief density of the random set $\bf X$ is
given by
 \beq
 f_{\bf X}({\bf B}) = \left. \frac{\delta \beta_{\bf X}({\bf S}) }{\delta {\bf B}} \right|_{{\bf
 S}=\emptyset}
 \eeq

Two useful rules of derivation are the following (see
also~\cite[p.\ 386 ff.]{mahlerbook}). Let $F$, $G$ be set
functions, and $a,b\in {\mathbb R}$. Then
 \beq\label{rule1}
 \frac{\delta (aF({\bf S})+bG({\bf S}))}{\delta {\bf B}}= a\frac{\delta F({\bf S})}{\delta {\bf B}}
 + b\frac{\delta G({\bf S})}{\delta {\bf B}}
 \eeq
 and
 \beq\label{rule2}
 \frac{\delta  F({\bf S})G({\bf S}) }{\delta {\bf B}}= \sum_{{\bf C}\subseteq {\bf B}}
 \frac{\delta  F ({\bf S})}{\delta {\bf B}} \frac{\delta  G({\bf S}) }{\delta ({\bf B} \setminus {\bf C})}
 \eeq

\subsection{Set integral} Let $f$ denote a set function defined by
 \[
 f({\bf X})=\left. \frac{\delta F ({\bf S})}{\delta {\bf X}}
 \right|_{{\bf S}=\emptyset}
 \]
 The set integral
of $f$ over the closed subset ${\bf S} \subseteq \mathbb S$ is
given by
 \begin{eqnarray}
 \lefteqn{ \int_{\bf S} f({ X}) \, \delta { X} }  \label{setintegral} \\
 &=&  f(\{ \emptyset \}) + \sum_{k=1}^\infty \frac{1}{k!}
 \int_{{\bf S}^k} f(\{ {\bf x}_1, \ldots,
 {\bf x}_k \}) d\bar{m}({\bf x}_1) \cdots d\bar{m}({\bf
 x}_k)  \nonumber
 \end{eqnarray}
where $f(\{ {\bf x}_1, \ldots,  {\bf x}_k \})=0$ if ${\bf x}_1,
\ldots, {\bf x}_k$ are not distinct. Since we are dealing with
{\em finite} random sets, the summation above contains only a
finite number of terms.

\subsection{Special case: $d=0$}
The special case $d=0$ (which corresponds to making $\mathbb S$ a
discrete finite set) reduces the set derivative to the {\em
M\"{o}bius inversion formula}~\cite[p.\ 43]{vihola}:
 \beq\label{mobius}
 f_{\bf X}({\bf A})= \sum_{{\bf B} \subseteq {\bf A}}
 (-1)^{| {\bf A} \setminus {\bf B}|} \beta_{\bf X}({\bf B})
 \eeq
and the set integral to
 \beq\label{setintegraldiscrete}
 f(\{ \emptyset \})+ \sum_{k=1}^\infty \frac{1}{k!}\sum_{ {\bf x}_1\neq \ldots \neq  {\bf x}_k }
 f(\{ {\bf x}_1, \ldots,  {\bf x}_k \})
   \eeq
where the summation is extended to all possible combinations of
$k$ distinct elements ${\bf x}_k \in {\bf S}$
 (in this case, the hybrid Lebesgue measure reduces to the
  counting measure, and hence the Lebesgue integrals in
  (\ref{setintegral}) become summations).

\subsection{Generalized fundamental theorem of calculus} Set
derivatives and set integrals turn out to be the inverse of each
other: we have
 \beq
 f({\bf X})=\left. \frac{\delta F ({\bf S})}{\delta {\bf X}}\right|_{{\bf S}=\emptyset} \Longleftrightarrow
 F({\bf S}) = \int_{\bf S} f({\bf X}) \, \delta {\bf X}
 \eeq
 By using the above result, belief functions and belief densities
 can be derived from one another.

\section*{Acknowledgments}
A preliminary version of this paper was presented at {\em WPMC
2006}, San Diego, CA, September 17--20, 2006.
 Ezio Biglieri was exposed to Random-Set Theory during his
visit to the Institute for Infocomm Research, Singapore, in 2005.
Discussions he had there with Professor Jiankang Wang are
gratefully acknowledged. He also benefited from discussions with
Professor Kung Yao, whose competence in target-tracking problems
helped him to put in the right perspective the topics discussed in
this paper. The authors are grateful to Shlomo Shamai (Shitz), who
called their attention to~\cite{plotnik}, which contains an
information-theoretical analysis of communication systems with a a
random number of active users, and to Daniele Angelosante for his
help with the derivation of numerical results.

{
The constructive comments of the reviewers, which
helped to improve the focus and clarity of the manuscript, are
gratefully acknowledged.}



\end{document}